\documentclass[aps,prf,a4paper,groupedaddress]{revtex4-2}
\usepackage[utf8]{inputenc} 
\usepackage[lmargin=2cm,rmargin=2cm,tmargin=2cm,bmargin=2cm]{geometry}
\usepackage{natbib}
\usepackage{subfig}
\usepackage{graphicx} 
\usepackage[svgnames,dvipsnames]{xcolor}  
\usepackage{fancyhdr}
\usepackage{fancybox}
\usepackage[labelfont=bf]{caption} 
\usepackage[T1]{fontenc} 
\usepackage{comment}
\usepackage{tikz}
\usetikzlibrary{shapes.geometric,calc}
\usetikzlibrary{shapes}
\usepackage{pgfplots} 

\usepackage{amsmath,amsfonts,amssymb}
\usepackage{float} 

\usepackage{colortbl} 
\usepackage[colorlinks=false,linkcolor=blue ,filecolor=red ]{hyperref}

 \fancypagestyle{plain}

\newcommand\cchi{\raisebox{1.5pt}{$\chi$}}

\begin{document}

\title{Hydrodynamic torque on a steadily rotating slender cylinder} 

\author{\small Jean-Lou Pierson, Mohammed Kharrouba, Jacques Magnaudet}

\date{\today}

\maketitle 

\section*{abstract}

Using fully-resolved simulations, we investigate the torque experienced by a finite-length circular cylinder rotating steadily perpendicularly to its symmetry axis. The aspect ratio $\cchi$, \textit{i.e.} the ratio of the length of
the cylinder to its diameter, is varied from 1 to 15. In the creeping-flow regime, we employ the slender-body theory to derive the expression of the torque up to order 4 with respect to the small parameter $1/\ln(2\cchi)$. Numerical results agree well with the corresponding predictions for $\cchi\gtrsim3$. We introduce an \textit{ad hoc} modification in the theoretical prediction to fit the numerical results obtained with shorter cylinders, and a second modification to account for the increase of the torque resulting from finite inertial effects. 
In strongly inertial regimes, a prominent wake pattern made of two pairs of counter-rotating vortices takes place. Nevertheless the flow remains stationary and exhibits two distinct symmetries, one of which implies that the contributions to the torque arising from the two cylinder ends are identical.  
We build separate empirical formulas for the contributions of pressure and viscous stress to the torque provided by the lateral surface and the cylinder ends. We show that, in each contribution, the dominant scaling law may be inferred from simple physical arguments. This approach eventually results in an empirical formula for the rotation-induced torque valid throughout the range of inertial regimes and aspect ratios considered in the simulations.







\section{Introduction}
\label{sec:intro}

Fibers and cylindrical rod-like particles are involved in numerous industrial processes such as paper making and food processing. In the chemical engineering industry, cylindrical pellets are used for instance for biomass extraction and oil refinement. 
Ice crystals growing and sedimenting in clouds also exhibit slender shapes and may be considered,  to a first approximation, as cylindrical rods. Hydrodynamic forces and torques acting on cylindrical particles generally depend critically on their orientation with respect to the relative incoming flow, which greatly complicates their prediction. For instance, in the creeping-flow approximation, the hydrodynamic force acting on a long isolated fiber moving broadside on is known to be twice as large as the force it experiences when moving along its axis \citep{batchelor1970}. This anisotropic behavior may have critical consequences in industrial processes in which pressure losses have to be reduced to a minimum. It is thus of primary importance to predict accurately the instantaneous orientation of the particles, and consequently the hydrodynamic torque acting on them. In dilute flow regimes, the time rate-of-change of the angular velocity is in most cases sufficiently small for the rotation-induced torque to balance the inertial torque due to the body inclination \cite{khayat1989}, possibly supplemented with a shear-induced inertial torque \citep{subramanian2005,einarsson2015}. 
This quasi-steady assumption \citep{cox1965} has proven accurate for predicting the motion of fibers settling in a vortical flow \citep{lopez2017} or in a fluid at rest at infinity \citep{roy2019}. However, a proper prediction of the angular velocity, hence of the instantaneous body orientation, under quite general conditions requires the influence of finite-length and inertial effects on the rotation-induced torque to be accurately quantified. 
With the aim of contributing to the modeling of this aspect for 
dilute suspensions, the present paper focuses on the rotation-induced torque acting on an isolated finite-length circular cylinder, from creeping-flow conditions to strongly inertial regimes. Indeed, the size of fibers encountered in applications covers a broad range corresponding to widely different flow regimes. While those employed in paper making industry have diameters typically in the range $15-30\,\mu$m and are a few millimeters long, typical rod-like catalysts involved in fluidized beds have diameters of roughly $1\,$mm and are $5$ to $10\,$mm long. In the former case, the Reynolds number based on the particle diameter is usually small, but the length-to-diameter aspect ratio is large. Conversely,  flow conditions relevant to rod-like catalysts usually correspond to moderate-to-large diameter-based Reynolds numbers and to aspect ratios of a few units.




The torque on a rotating sphere was investigated numerically by \citet{dennis1980}. The results were found to agree well with Kirchhoff's solution \cite{kirchhoff1876} at low Reynolds number. At large Reynolds number, the solutions were found to tend toward the predictions of boundary layer theory. An empirical law for the torque, bridging the two limits, was proposed. Although a large body of literature if available regarding the forces experienced by non-spherical particles (see for instance \cite{loth2008} for a review), much less is known regarding the rotation-induced torque. In the creeping-flow limit, \citet{batchelor1970} made use of the slender-body theory to evaluate the force and torque acting on long cylindrical bodies.  In the opposite limit of nearly-inviscid flows, \citet{kry1974,kry1974b} considered the torques experienced by rotating oblate spheroids. They provided a range of applicability of the quasi-steady approximation for the aerodynamic torque. In particular, they showed that a spinning motion about the spheroid minor axis modifies the boundary layer but does not affect significantly the pressure distribution past the body.



In this work, we determine numerically the torque experienced by a steadily rotating circular cylinder of finite length, from creeping-flow conditions to strongly inertial regimes. We briefly present the problem and the computational strategy in Sec. \ref{sec:problem}. In Sec. \ref{sec:stokes}, we first compute the torque in the low-Reynolds regime and compare the numerical results with the prediction of the slender-body approximation, which we extend to fourth order in appendix \ref{app:slender}. 
 Then we examine the influence of finite inertial effects and incorporate an empirical correction in the creeping-flow formula to take this influence into account. Moderate-to-large Reynolds number flow conditions are examined in Sec. \ref{sec:inertia}. In the spirit of our recent investigation on the loads acting on an inclined translating cylinder \citep{kharrouba2020}, practical estimates for the torque acting on the body are derived from the simulations by considering separately the contribution of viscous stresses and pressure on each part of the body, \textit{i.e.} the lateral surface and the two flat ends. We summarize our results in Sec. \ref{sec:discussion} and discuss their implications with respect to the variations of the body rotation rate with the aspect ratio and the Reynolds number.

\section{Problem definition and numerical approach}
\label{sec:problem}

We consider the flow induced by a circular cylinder of length $L$ and diameter $D$ rotating about an axis perpendicular to its symmetry axis and passing through its center of inertia. The fluid is Newtonian, with density $\rho$ and dynamic viscosity $\mu$, and is at rest at infinity. The problem depends on two dimensionless parameters, the aspect ratio $\cchi = L/D$ and the Reynolds number which we define as $\text{Re} = \rho \Omega L D / (2\mu) $. This definition assumes that $D$ is the relevant length scale of the flow, while the characteristic velocity $U$ is assumed to be $\Omega L / 2$. In what follows, we investigate the flow and torque induced by the cylinder rotation in the range $ 0.05 \leq \text{Re} \leq 240 $ and $ 1 \leq \cchi \leq 15$. In inertia-dominated regimes, we only consider aspect ratios in the range $ 2 \leq \cchi \leq 8$ to reduce the computational cost.

The computational strategy is based on the formulation introduced by \citet{mougin2002} and we refer to the original article for details. In short, the Navier-Stokes equations are solved for the \textit{absolute} velocity field $\mathbf{u}$ (the one measured by a fixed observer) using a coordinate system rotating and possibly translating with the body. Hence, the body is fixed and the fluid rotates about it. Note that this formulation differs from the classical one involving the Coriolis pseudo-force, since  the latter makes use of the same coordinate system but considers the \textit{relative} velocity field. 
 In the present approach, the fluid is at rest at infinity and obeys the no-slip condition $\mathbf{u}=\boldsymbol{\Omega}\times\mathbf{x}$ at the cylinder surface, $\mathbf{x}$ denoting the local position from the body centroid.\\
The computational domain is a large cylinder with the same axis as the body. Its diameter is equal to its length. The discretization of the fluid domain is mostly similar to that used by \citet{kharrouba2020} but the configuration of interest here introduces some specificities. For low-to-moderate Reynolds numbers, the grid is uniform throughout the fluid region extending up to $0.5D$ from the body surface, with cell sizes ranging from $D/16$ to $D/20$ depending on $\text{Re}$. Beyond this region, the cell size increases with the distance to the body following a geometric law with a common ratio close to $1.07$. $32$ cells are uniformly distributed in the azimuthal direction. The length and diameter of the domain range from $60D$ for $\cchi=1$ and $2$ to $215D$ for $\cchi=10$ and $15$. Such large dimensions are required due to to the slow decrease of the disturbance induced by the body rotation. On the outer surface of the domain, the normal component of $\mathbf{u}$ is assumed to be zero, together with the normal derivative of the tangential components. We select this `free-slip' condition rather than a non-reflecting outlet condition because the tiny remaining velocity disturbance has an inward component on some parts of the outer surface, which could create numerical instabilities. In the inertia-dominated regime, detailed tests showed that the boundary layer, whose thickness is estimated to be $\delta \sim D / Re^{1/2}$ on the body ends and on the part of the lateral surface close to them, is accurately captured with 6 cells. Hence, the cell size in the fluid region extending up to $0.5D$ from the body surface is set to $\delta/6$. $64$ cells are uniformly distributed in the azimuthal direction. The domain size is set to $L+30D$, \textit{i.e.} the outer boundary is located $15D$ apart from the cylinder ends. Since the velocity disturbance decays much faster with the distance to the body than in the low-to-moderate $\text{Re}$ regime, a non-reflecting boundary condition \cite{magnaudet1995} is used on the outer boundary. 

\section{From the creeping-flow regime to moderate Reynolds numbers}
\label{sec:stokes}
For $\text{Re}=0$, the slender-body theory provides a convenient framework to estimate the torque on a rotating finite-length cylinder as long as the aspect ratio is much larger than unity. \citet{batchelor1970} carried out the third-order expansion with respect to the small parameter $\epsilon = 1/\ln (2 \cchi)$ for this case, computing the corresponding coefficients numerically. In appendix \ref{app:slender} we derive these coefficients analytically up to order 4, based on the iterative technique developed by \citet{keller1976}.    

\begin{figure}[h]
\centering
\includegraphics[width=7cm]{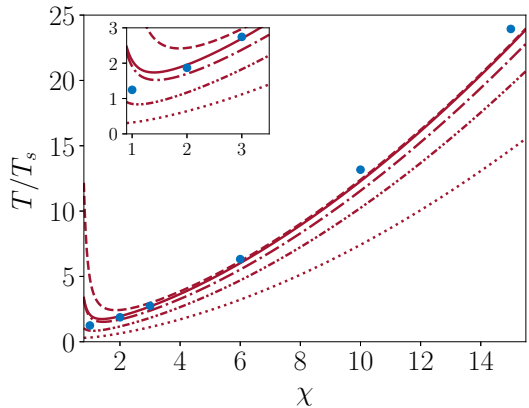}
\caption{Torque on a rotating finite-length cylinder, normalized by the torque on a rotating sphere with the same volume. Dotted, dash-dotted-dotted, dash-dotted and dashed lines: predictions of $1^{rth}$-, $2^{nd}$-, $3^{rd}$- and $4^{th}$-order slender-body approximations, respectively; solid line: semiempirical formula (\ref{eq:sl4_bism}), $\bullet$: numerical results for $Re=0.05$.}
\label{fig:stokes}
\end{figure}

\noindent The corresponding results are displayed in Fig. \ref{fig:stokes}. The torque is made dimensionless by dividing it by its counterpart on a rotating sphere with the same volume, namely $T_s = -\pi \mu \Omega \mathcal{D}^3 $, with $\mathcal{D} =(\frac{3}{2} \cchi)^{1/3}D$. The torque on a rotating cylinder is larger than that on the equivalent sphere for all aspect ratios $\cchi\gtrsim1$. Indeed, at leading order, the ratio of the two torques varies as $\cchi ^2 / \log (2 \cchi)$. Not surprisingly, the first and second-order approximations do not match the numerical results well, even for large aspect ratios. The third-order approximation provides a better estimate but still under-predicts the actual torque. A significantly better agreement is obtained with the fourth-order approximation for $\cchi \geq 3$. All approximations diverge as $\cchi\rightarrow1/2$, making the agreement deteriorate for $\cchi\lesssim3$. However, the third-order approximation is still accurate in the range $1.5\lesssim\cchi\lesssim3$. Making use of these observations, we introduce an \textit{ad hoc} modification of the original expansion.  That is, we multiply the fourth-order term by a function of $\cchi$ that quickly tends towards unity as $\cchi$ increases and towards zero when $\cchi \rightarrow 1/2$, and varies in such a way that the behavior of the third-order expansion is recovered in the range $2\lesssim\cchi\lesssim3$. The full expression reads     
\begin{align}
T=-&\frac{\pi\mu\Omega L^3}{3}\left[\epsilon+\epsilon^2\left(\frac{11}{6}-\ln 2\right)+\epsilon^3\left(\frac{161}{36}- \frac{\pi ^2}{12}-\frac{11}{3}\ln 2+(\ln 2)^2\right)\right. \nonumber\\
                             &\left. +\epsilon^4\left(1-\frac{1}{(2\cchi)^{1.2}}\right)^5 \left(- \frac{5}{4}\zeta (3)+\frac{1033}{72}-\ln ^3 (2)+\frac{11}{2}\ln ^2(2) - \frac{161}{12}\ln 2-\pi ^2 \left(\frac{11}{24}-\frac{1}{8} \ln (4)\right)\right)\right]\,.
\label{eq:sl4_bism}
\end{align}
The formula (\ref{eq:sl4_bism}) agrees well with the numerical results for $\cchi \gtrsim 2$. It slight deviates from the numerical result for $\cchi \geq 10$. This is not unlikely since the Reynolds number based on $L$ instead of $D$ is of $\mathcal{O}(1)$ in this case, suggesting that inertial effects are already significant. 
The inset in Fig. \ref{fig:stokes} indicates that the normalized torque on a cylinder with $\cchi=1$ is approximately $1.25$, \textit{i.e.} the torque is larger than that on the equivalent sphere, $\vert T_s\vert\approx1.145\pi\mu D^3\Omega$. In appendix \ref{app:min}, we show how bounds for the torque may be derived from the minimum dissipation theorem. For $\cchi=1$, these predictions indicate that the torque is such that $\pi\mu D^3\Omega \leq \vert T\vert \leq  2^{3/2}\pi\mu D^3\Omega$. The numerical result obviously stands in the allowed interval. 


\begin{figure}[h]
\centering
\includegraphics[width=5cm]{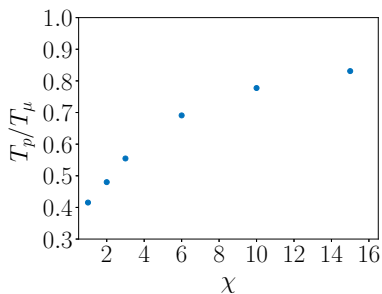}
\caption{Ratio of the pressure-to-shear stress contributions to the torque for $Re=0.05$.}
\label{fig:contrib}
\end{figure}

Figure \ref{fig:contrib} shows how the ratio $T_p/T_\mu$ between the pressure and shear stress contributions to the torque vary with the aspect ratio. The former contribution is observed to be smaller than the latter whatever $\cchi$. Keeping in mind that $T_p$ would be zero for a sphere, it is no surprise that $T_p/T_\mu$ increases with $\cchi$. This ratio is slightly larger than $0.8$ for $\cchi = 15$ and the observed variation suggests that it becomes independent of $\cchi$ and of $\mathcal{O}(1)$ for long enough cylinders. 


\begin{figure}[h]
\centering
\includegraphics[width=5cm]{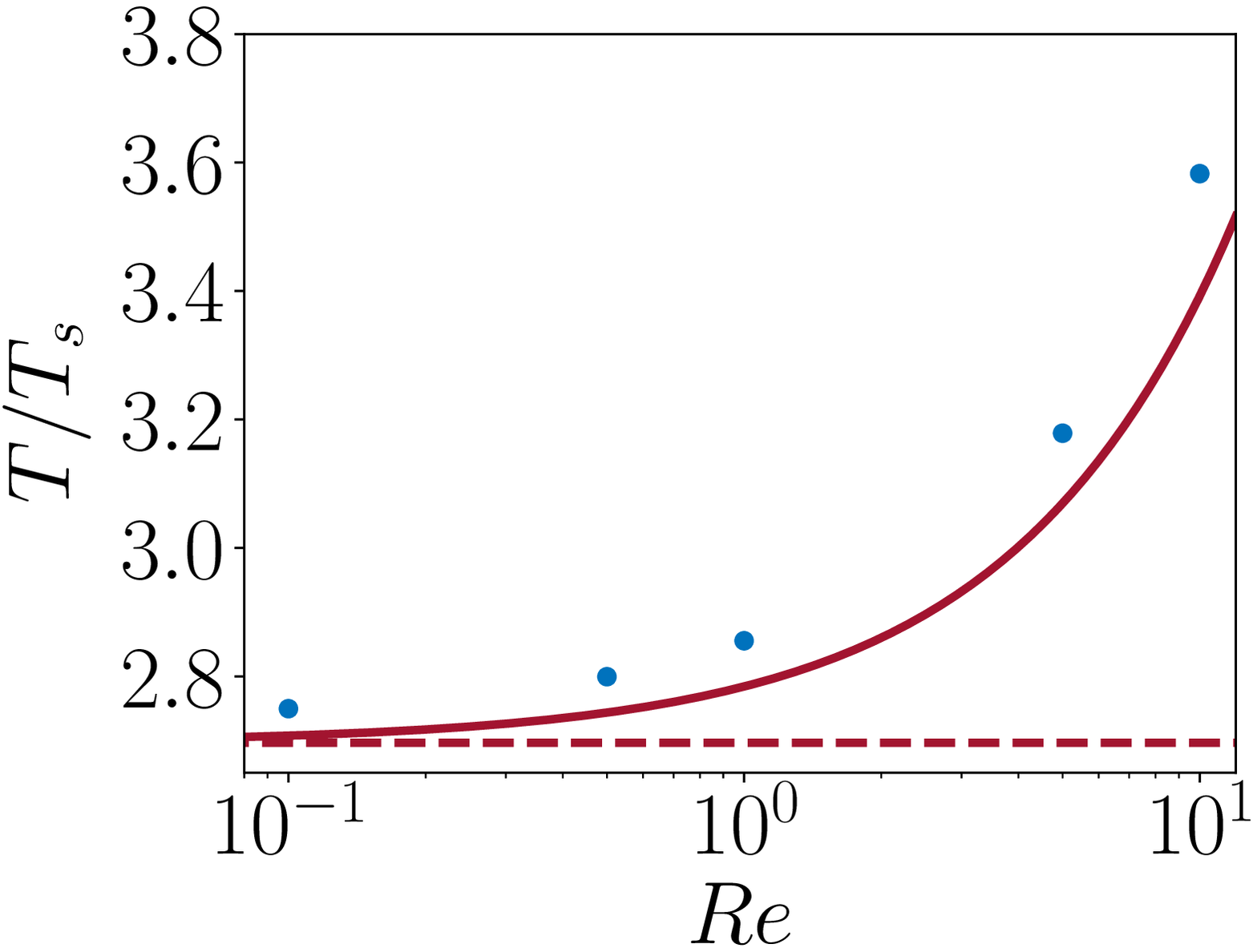} \quad \includegraphics[width=5.15cm]{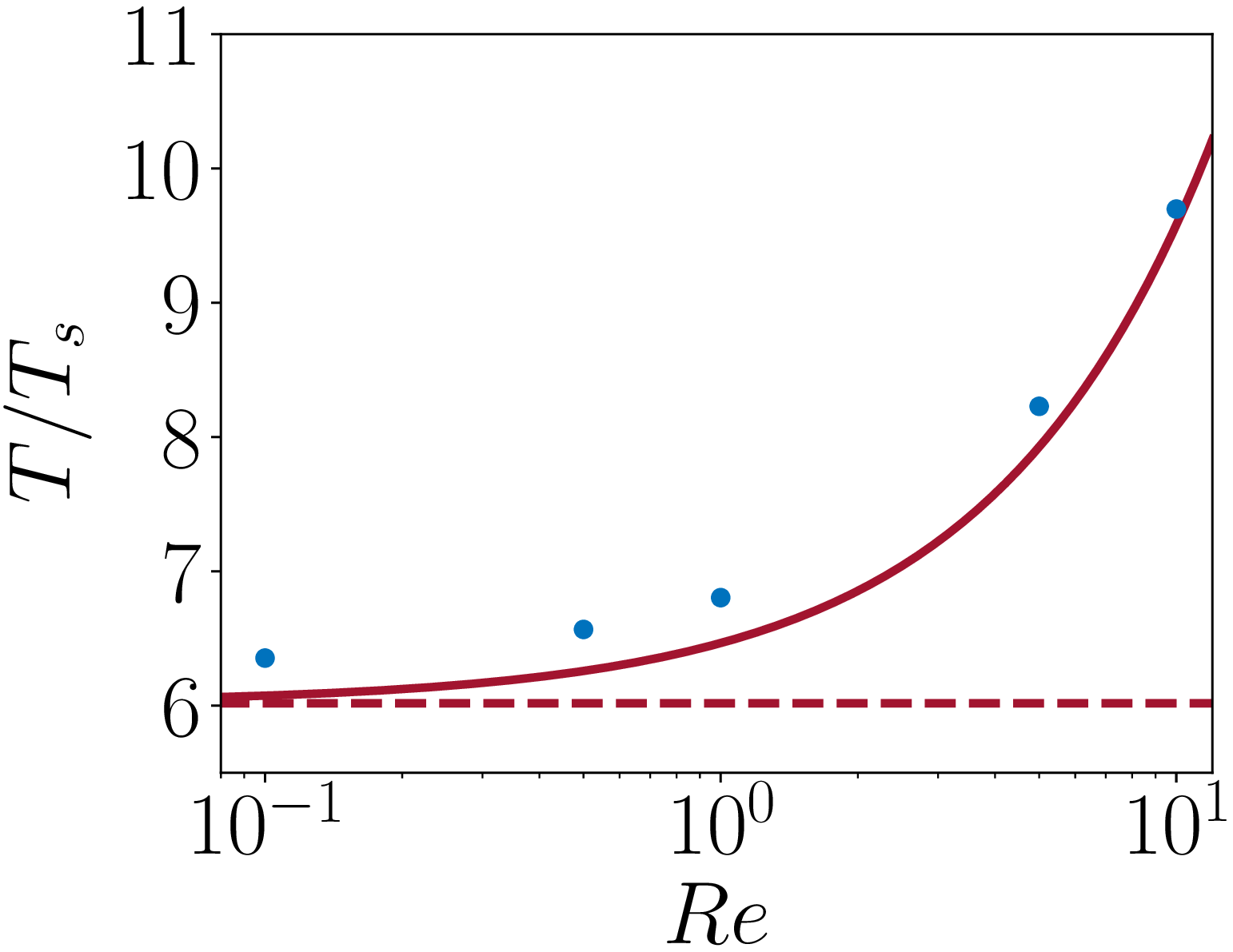} \quad \includegraphics[width=4.9cm]{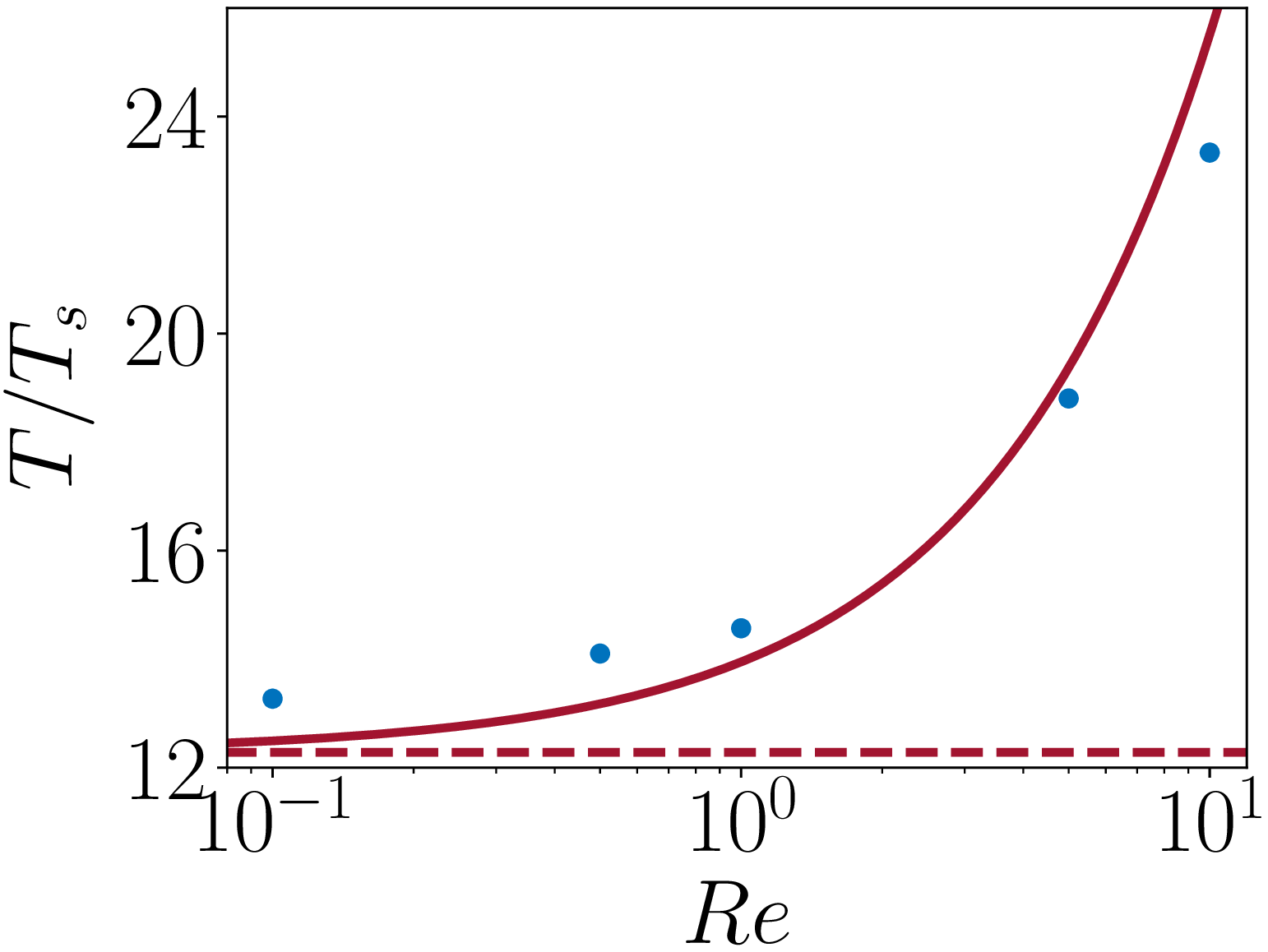}\\
\vspace{-4mm}
\hspace{-25mm}(a) \hspace{4.9cm} (b) \hspace{4.9cm} (c)\\
	\vspace{-35mm}
	\hspace{-13mm}$\cchi=3$\hspace{48mm}$\cchi=6$\hspace{46mm}$\cchi=10$
	\vspace{32mm}\\
\caption{Normalized torque as a function of the Reynolds number in the range $0.1\leq\text{Re}\leq10$. $\bullet$: numerical results; dashed line: creeping-flow prediction (\ref{eq:sl4_bism}); solid line: semiempirical formula (\ref{eq:sl4_bism_re}). The slight difference between numerical results and predictions of (\ref{eq:sl4_bism}) for $Re \ll 1$ is already present in figure 1 but is magnified here, owing to the chosen origin of the vertical axis.}
\label{fig:T_moderate}
\end{figure}

Figure \ref{fig:T_moderate} displays the increase of the normalized torque as function of $\text{Re}$ for three aspect ratios and Reynolds numbers up to $10$. In the range $0.05 \leq \text{Re} \leq 1$, the torque is seen to vary by less than 5\% for $\cchi =3$, while it varies by more than $10\%$ for $\cchi=10$. For $\text{Re} =10$, the torque on the longest cylinder has doubled with respect to its value in the creeping flow regime, while it has only increased by nearly $40\%$ for $\cchi =3$. That the relative contribution of inertial effects to the torque increases with the aspect ratio is no surprise. Indeed, the relevant characteristic length scale of the flow is the lever arm $L/2=\cchi D/2$ rather than $D$, so that the relevant ratio of inertial to viscous effects is $\frac{\cchi}{2}\text{Re}$. 
Finite inertial effects acting on a slender body inclined with respect to a uniform incoming flow were considered by \citet{khayat1989}, using the method of matched asymptotic expansions. Their predictions, in which inertial corrections affect the $\mathcal{O}(\epsilon^2)$-terms of the expansion, quantify the inertia-induced increase of the drag and lift forces. In the case of a cylinder (more generally a body with a straight centerline), these predictions also reveal the existence of a a nonzero inertial torque which tends to rotate the body broadside on with respect to the incoming flow. To the best of our knowledge, finite-inertia effects have not been considered for a slender body rotating in a fluid at rest at infinity. Adapting the approach of \cite{khayat1989} to this configuration is beyond the scope of the present work. Instead, we only attempted to use the numerical results to extend the formula (\ref{eq:sl4_bism}) empirically to $\mathcal{O}(1)$-Reynolds numbers by suitably altering the pre-factor of the $\mathcal{O}(\epsilon^2)$-term.  Based on the numerical findings and the above remarks, we modified (\ref{eq:sl4_bism}) in the form
\begin{align}
T=&\frac{-\pi\mu\Omega L^3}{3}\left[\epsilon+\epsilon^2\left(\frac{11}{6}-\ln 2 + f(\cchi,\text{Re})\right)+\epsilon^3\left(\frac{161}{36}- \frac{\pi ^2}{12}-\frac{11}{3}\ln 2+(\ln 2)^2\right)\right. \nonumber\\
                             &\left. +\epsilon^4\left(1-\frac{1}{(2\cchi)^{1.2}}\right)^5 \left(- \frac{5}{4}\zeta (3)+\frac{1033}{72}-\ln ^3 (2)+\frac{11}{2}\ln ^2(2) - \frac{161}{12}\ln 2-\pi ^2 \left(\frac{11}{24}-\frac{1}{4} \ln 2\right)\right)\right],
\label{eq:sl4_bism_re}
\end{align}

\noindent with $f(\cchi,\text{Re}) = 0.0337\cchi^{1.3}\text{Re}^{0.9}$. Only the second-order term has been modified since, according to Fig. \ref{fig:T_moderate}, this change appears to be sufficient to capture most of the finite-$\text{Re}$ variations of the torque up to $\text{Re}=10$.



\section{Flow stucture and torque in inertia-dominated regimes}
\label{sec:inertia}

\begin{figure}[h]
\centering
\includegraphics[height=4cm]{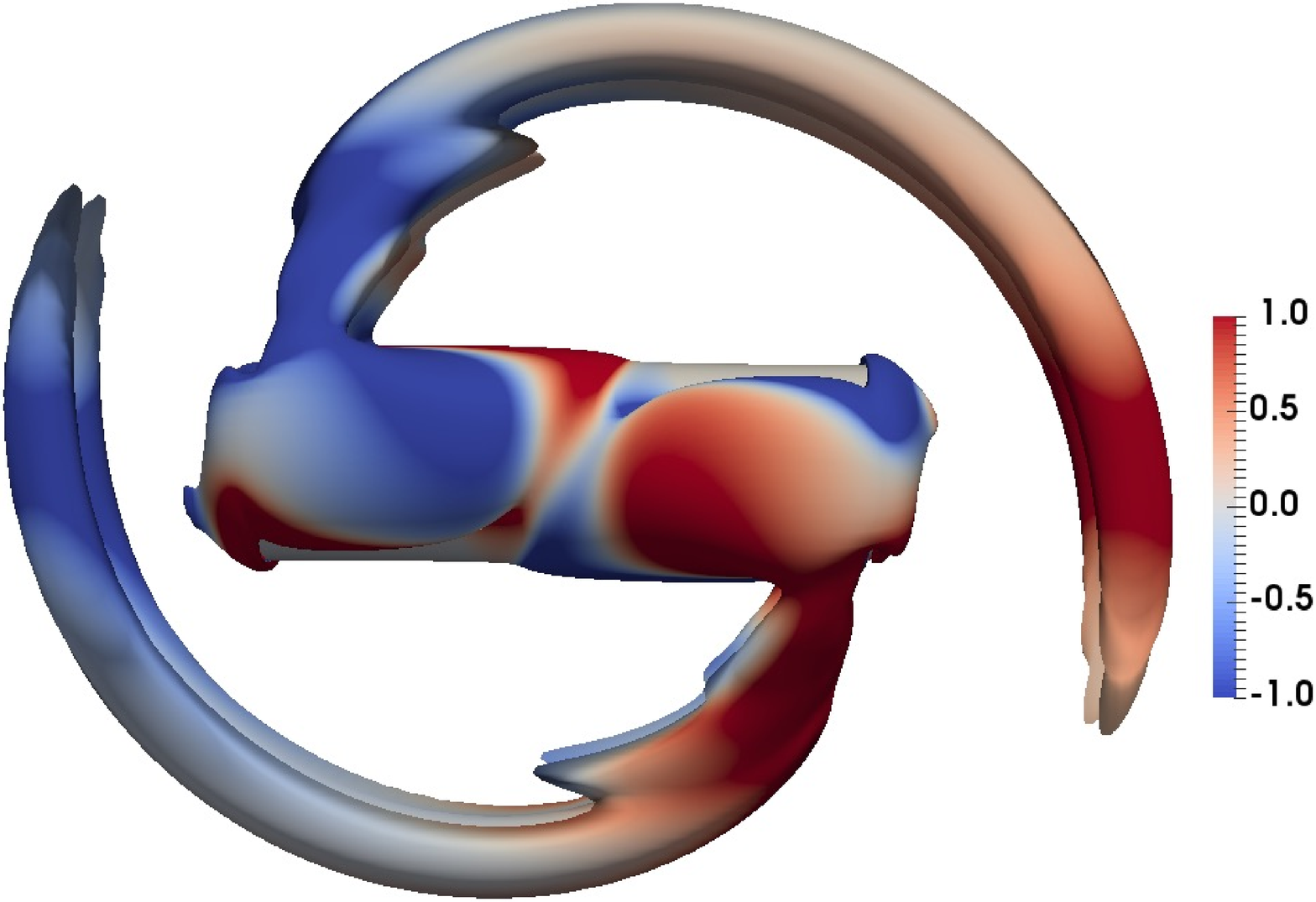} \quad \includegraphics[height=4.5cm]{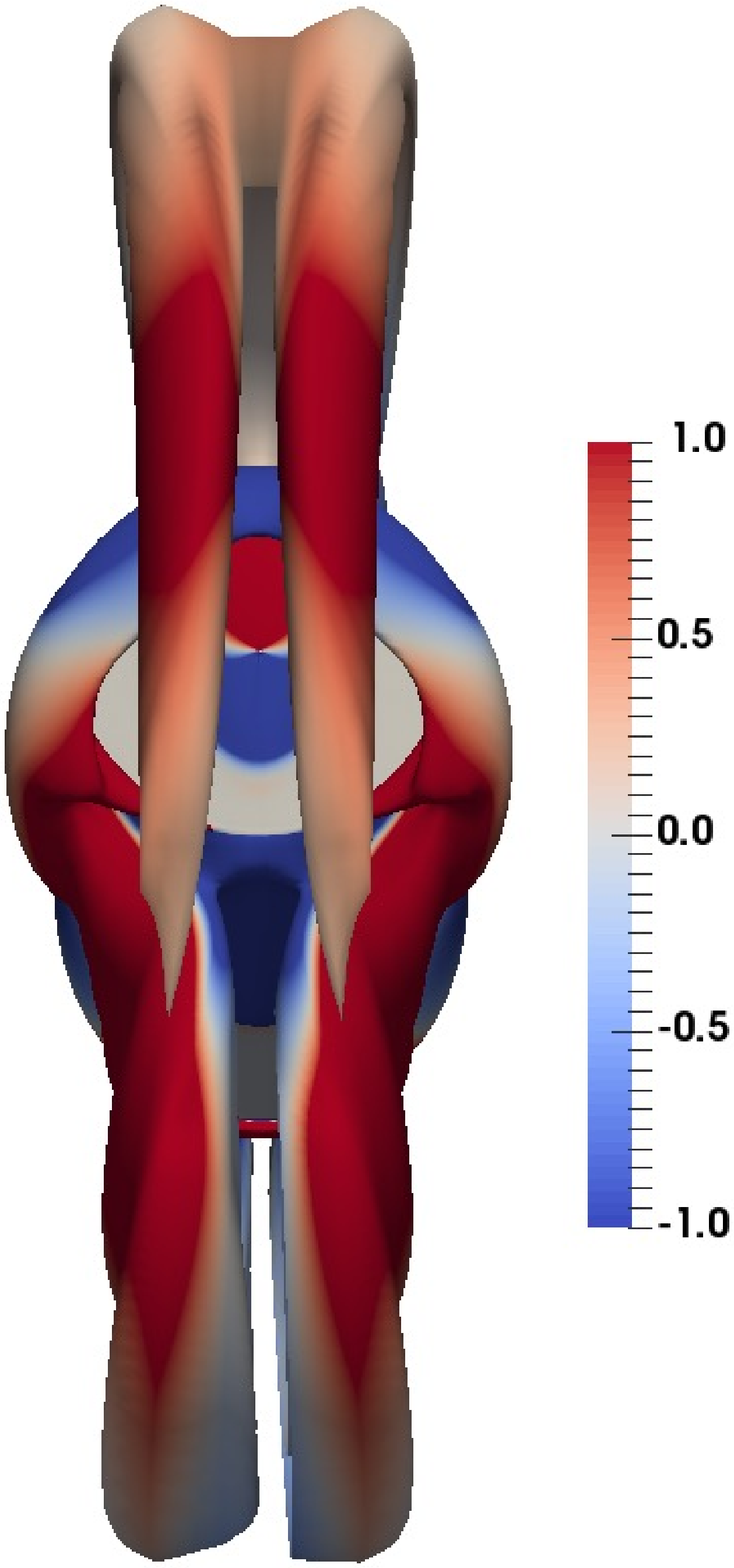}\\
\hspace{0.5cm}(a) \hspace{5cm} (b)
\caption{Wake past a rotating cylinder with $\cchi = 3$ for $\text{Re} = 220$. (a): side view from a fixed $z$-plane (see Fig. \ref{fig:scheme} for the definition of the coordinate system); (b): front view from a fixed $x$-plane. The wake is visualized using the $Q$-criterion \cite{hunt1988}. 
The $0.001$ iso-surface of $Q$ is colored with the magnitude of the normalized transverse vorticity $\omega_z/\Omega$.}
\label{fig:vortex_pair}
\end{figure}

Inertia-dominated regimes with Reynolds numbers beyond those considered before are characterized by the presence of a prominent wake. As Fig. \ref{fig:vortex_pair} shows, two pairs of counter-rotating vortices take place. Each pair emanates from one end of the cylinder and bends downstream under the effect of the flow rotation, resulting in helical vortices similar to those observed past rotating blades. Since the strength of the vorticity generated at the cylinder surface increases with $\text{Re}$, so does the length of the vortices. No unsteadiness, \textit{i. e.} no vortex shedding, was observed throughout the whole range of Reynolds number considered here ($\text{Re} \leq 240$). In this regime, the wake exhibits two distinct symmetries. As Fig. \ref{fig:vortex_pair}(b) reveals, the plane perpendicular to the rotation axis and passing through the cylinder centroid is a symmetry plane. This planar symmetry, combined with the flow steadiness, implies that all force components are zero at any instant of time and that the torque is collinear with the rotation axis. 


\begin{figure}[h]
\centering
\begin{tikzpicture}[scale=1.]
\fill[color=orange!50, thick] (-2,0) rectangle +(0.5,0.755) ;
\fill[color=orange!50, thick] (-2,0.75) rectangle +(2,0.5) ;

\fill[color=orange!50, thick, rotate=180] (-2,0) rectangle +(0.5,0.755) ;
\fill[color=orange!50, thick, rotate=180] (-2,0.75) rectangle +(2,0.5) ;

\fill[color=cyan!50, thick] (-2,0) rectangle +(0.5,-0.755) ;
\fill[color=cyan!50, thick] (-2,-0.75) rectangle +(2,-0.5) ;

\fill[color=cyan!50, thick, rotate=180] (-2,0) rectangle +(0.5,-0.755) ;
\fill[color=cyan!50, thick, rotate=180] (-2,-0.75) rectangle +(2,-0.5) ;

\node [cylinder, rotate=0, draw, minimum height=3.2cm, minimum width=1.5cm]  at (-0.1, 0, 0) (c) {}; 
\draw[dashed] (-1.5,-0.75) rectangle +(3,1.5);

\draw[->, thick] (0,0)--(1,0) node[right]{$x$};
\draw[->, thick] (0,0)--(0,1) node[above]{$y$};
\draw[->, thick] (0,0,0)--(0,0,1.5) node[above]{$z$};
\draw[->] (0.25,0,0.25) arc (-10:350:0.2)node[below]{$\Omega$}; 



\foreach \y in{ 1.5, 1.6,...,2.5 }
\draw[color=black,->] (\y,0,0) -- (\y,{1/\y^1.3},0) ;
\draw[dashed] (1.5,0)--(2.5,0);
\draw[color=black] (2.2,0.6) node[right]{$u_y(x)$};

\foreach \y in{ -2.5,-2.4,...,-1.5}
\draw[color=black,->] (\y,0,0) -- (\y,{1/\y^1.3},0) ;
\draw[dashed] (-1.5,0)--(-2.5,0);
\draw[color=black] (-3.1,-0.6) node[right]{$u_y(x)$};

\end{tikzpicture}
\caption{Sketch of the flow configuration near the rotating cylinder. Colors help identify the flow symmetries.}
\label{fig:scheme}
\end{figure}
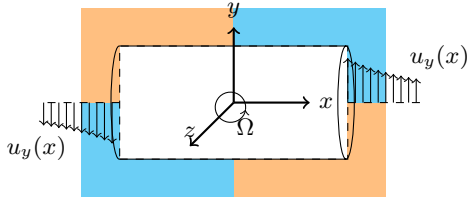

Figure \ref{fig:vortex_pair}(a) reveals a second symmetry resulting from the combination of two mirror symmetries. One is with respect to the plane containing both the rotation axis and the cylinder symmetry axis (defined as the $z$- and $x$-axes in Fig. \ref{fig:scheme}, respectively). The other is with respect to the plane orthogonal to the previous plane and containing again the rotation axis ($(y,z)$-plane in Fig. \ref{fig:scheme}). 
This symmetry is illustrated with colored areas in Fig. \ref{fig:scheme}. For instance, the pressure at a given point in the upper half of the right end (blue area) is identical to the pressure at the mirror point of the left end (blue area again). Since the cross product $\mathbf{x}\times\mathbf{n}$ (with $\mathbf{n}$ the unit normal to the cylinder pointing into the fluid) is also the same at the two locations, it turns out that the pressure contributions to the torque provided by the two ends are identical. The $x$-variation of the tangential velocity $u_y$ in the vicinity of both ends is schematized in Fig. \ref{fig:scheme}. As this sketch suggests, the same property holds true for the shear stress contribution  provided by both ends. \\

\begin{figure}[h]
\centering
\includegraphics[width=5cm]{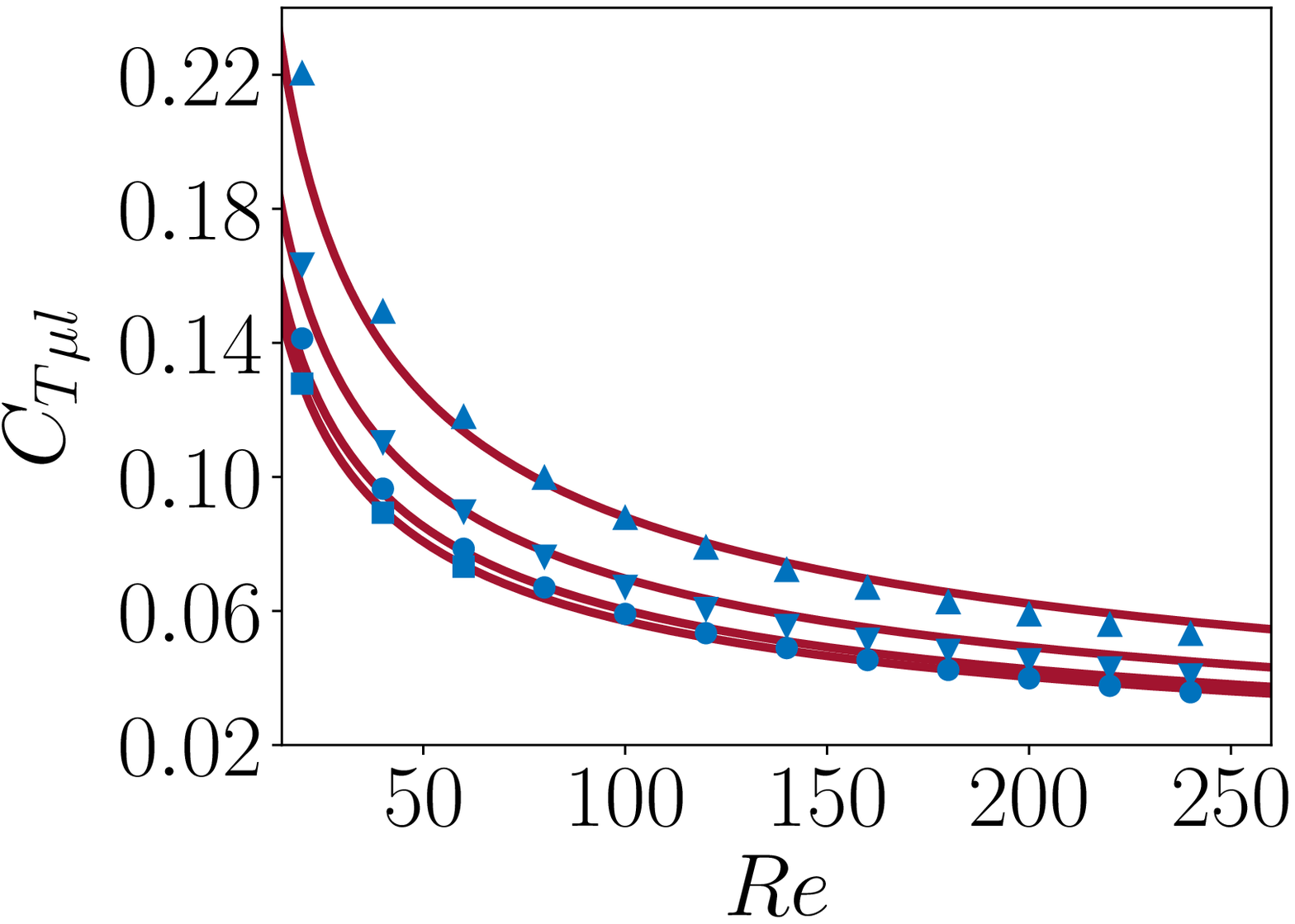} \quad \includegraphics[width=5cm]{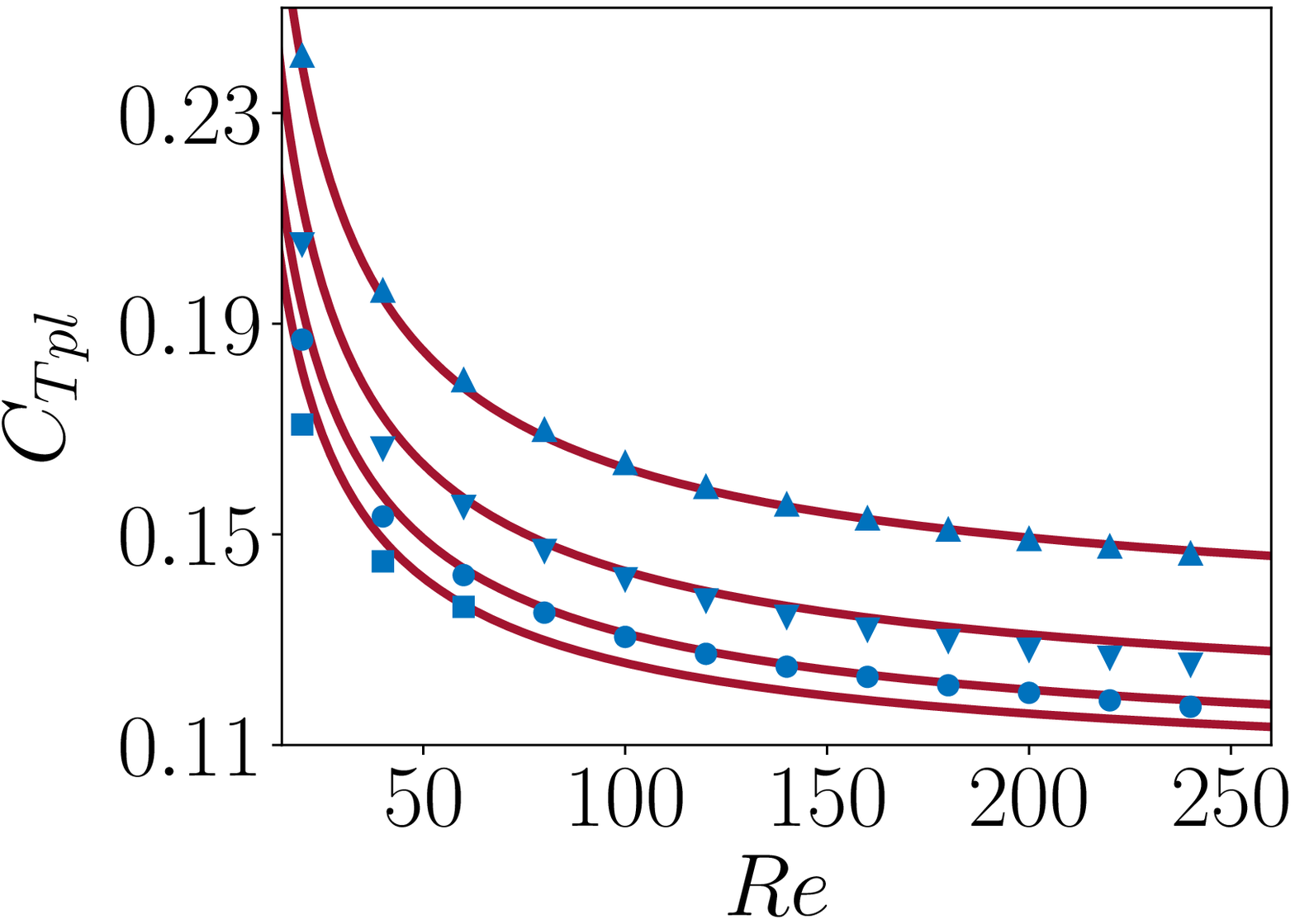} \\
\vspace{-3mm}
\hspace{-2.7cm}(a) \hspace{4.9cm} (b)
\caption{Contributions to the torque arising from the lateral surface as a function of $\text{Re}$. (a): viscous (shear stress) contribution; (b): pressure contribution. $\blacktriangle$: $\cchi =2$, $\blacktriangledown$: $\cchi =3$, $\bullet$: $\cchi =5$, $\blacksquare$: $\cchi =8$. Solid line: empirical fits (\ref{eq:Ctmul}) and (\ref{eq:Ctpl}).}
\label{fig:ctmu_p}
\end{figure}


Figure \ref{fig:ctmu_p} displays the contributions to the torque arising from the lateral surface. For $\cchi=8$, computations were only carried out up to $\text{Re} =60$, owing to the cost associated with the large grids required to capture the flow details at higher $\text{Re}$. From now on, we characterize the torque through the coefficient $C_T$ obtained by normalizing $T$ with the reference torque $\frac{1}{8}\rho \Omega ^2 L^4D$, since the characteristic velocity is $\Omega L/2$ and the characteristic surface is $LD$. For each aspect ratio, the viscous contribution (Fig. \ref{fig:ctmu_p}(a)) decreases approximatively as $\text{Re}^{-1/2}$, indicating that the magnitude of the shear stress is dictated by the boundary layer thickness. This contribution is also seen to depend only weakly on $\cchi$ as soon as $\cchi\gtrsim5$.  Hence, the corresponding coefficient can be fitted with the simple expression
\begin{equation}
C_{T\mu l} = (1.32\cchi^{-2}+0.55)\text{Re}^{-1/2}\,.
\label{eq:Ctmul}
\end{equation}

 \noindent The pressure contribution (Fig. \ref{fig:ctmu_p}(b)) decreases as $\cchi$ or $\text{Re}$ increases, gradually tending toward a constant value for large Reynolds numbers whatever $\cchi$. Numerical data are properly fitted with the three-term correlation
\begin{equation}
C_{Tpl}=1.21\cchi ^{-0.23} \text{Re} ^{-0.75}+0.12\cchi ^{-2}+0.1\,. 
\label{eq:Ctpl}
\end{equation}
This fit suggests that the pressure contribution to the torque tends toward $0.1$ for large enough Reynolds numbers and infinitely long cylinders. Although a torque coefficient independent of both $\cchi$ and $\text{Re}$ is to be expected in this limit, it must be kept in mind that only the steady regime is considered here, so that (\ref{eq:Ctpl}) may not be valid in the unsteady regimes that take place for sufficiently large $\text{Re}$. 

\begin{figure}[h]
\centering
\includegraphics[width=5cm]{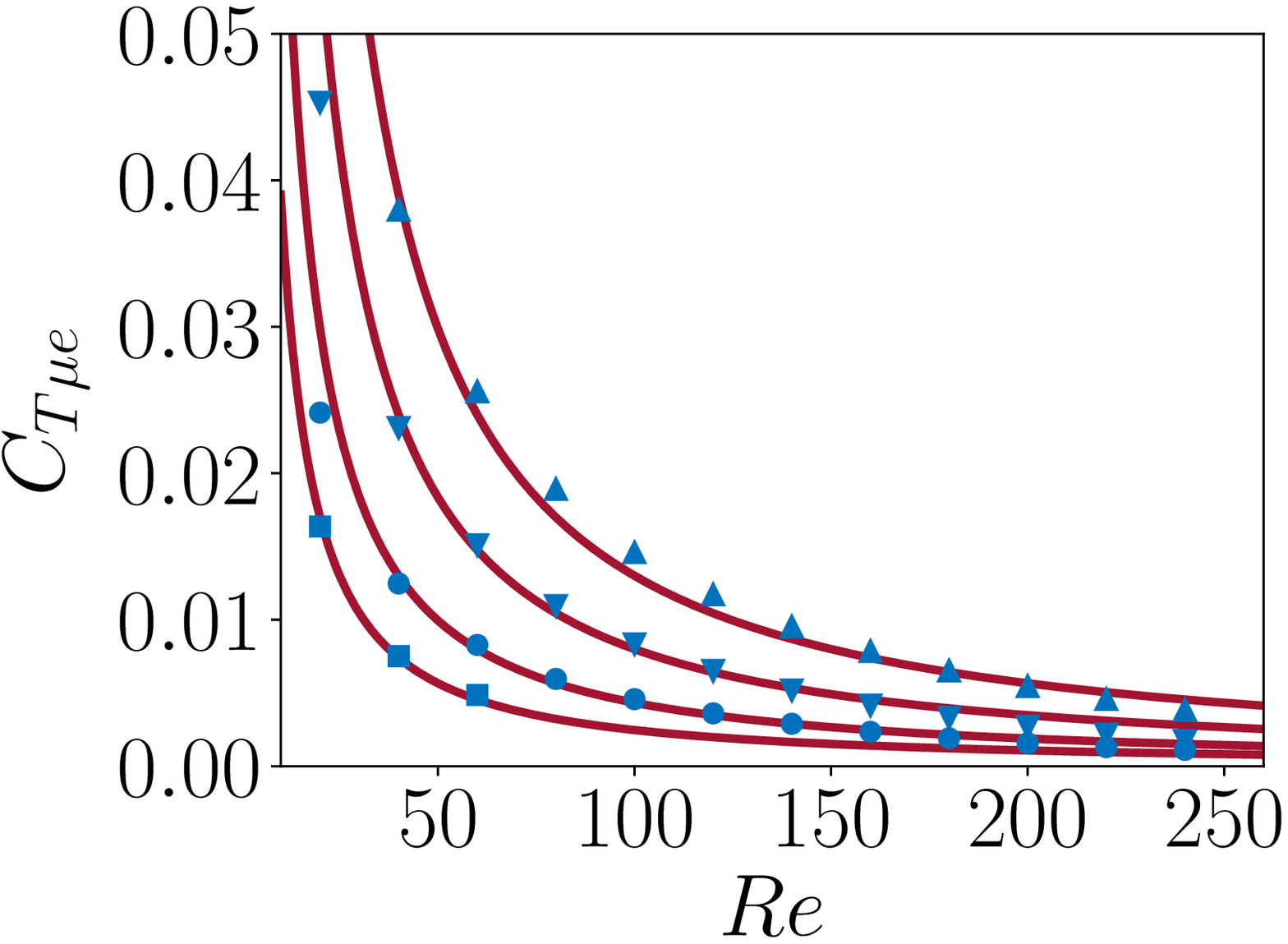} \quad \includegraphics[width=5cm]{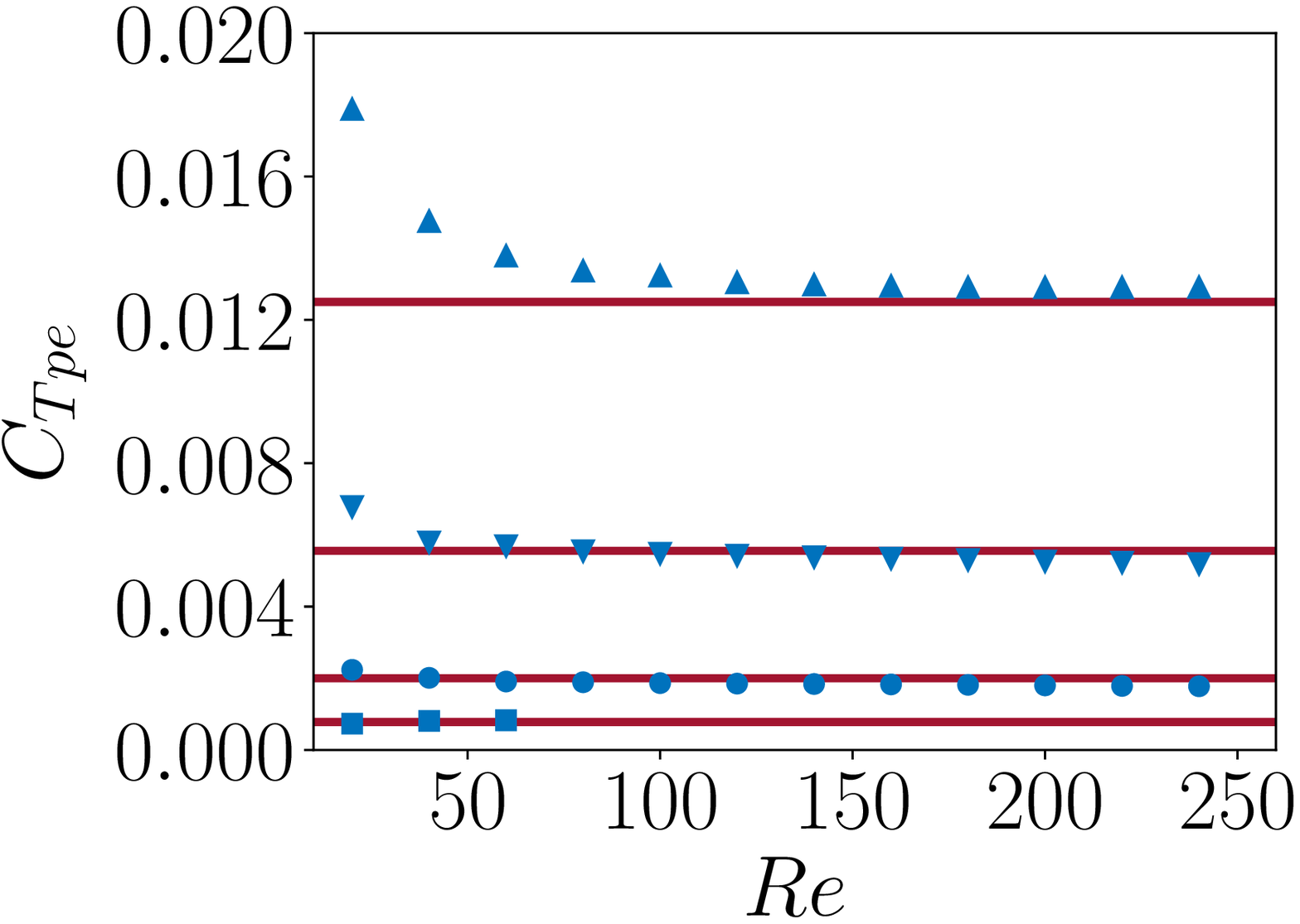}\\
\vspace{-3mm}
\hspace{-2.7cm}(a) \hspace{4.9cm} (b)
\caption{Contributions to the torque arising from each of the cylinder ends as a function of $\text{Re}$. (a): viscous (shear stress) contribution; (b): pressure contribution. $\blacktriangle$: $\cchi =2$, $\blacktriangledown$: $\cchi =3$, $\bullet$: $\cchi =5$, $\blacksquare$: $\cchi =8$. Solid line: empirical fits (\ref{eq:Ctmud}) and (\ref{eq:Ctpd}). 
}
\label{fig:ctmu_p_d}
\end{figure}
Figure \ref{fig:ctmu_p_d} displays the contribution of each cylinder end to the torque. The viscous contribution (Fig. \ref{fig:ctmu_p_d}(a)) decreases strongly as $\cchi$ or $\text{Re}$ increases. It is adequately fitted with the simple formula  
\begin{equation}
C_{T \mu e} = 7.5 \cchi ^{-1.2}\text{Re}^{-1.2}.
\label{eq:Ctmud}
\end{equation}


\noindent It is worth noting that the negative $\text{Re}$-exponent is significantly higher than that provided by the boundary layer theory. The reason is that the flow is massively separated in the end regions, and the typical length scale of the corresponding recirculation is $D$. For this reason, the magnitude of the shear stress on the cylinder ends is governed by the cylinder diameter, not by the boundary layer thickness. This scaling yields viscous stresses of $\mathcal{O}(\mu U/D)$, hence contributions to the torque of $\mathcal{O}(\mu ULD)$, which results in $C_{T\mu e} \sim Re^{-1}\cchi^{-1}$, close to the behavior synthesized by (\ref{eq:Ctmud}). \\
\indent  As Fig. \ref{fig:ctmu_p_d}(b) reveals, the pressure contribution to the torque arising from the cylinder ends does not vary significantly with the Reynolds number in the range of interest here, except for the shortest cylinder for which some decrease is observed for $\text{Re} \leq 80$. Compared to the pressure contribution provided by the lateral surface, $C_{Tpe}$ is one order of magnitude smaller for $\cchi=2$ and approximately $50$ times smaller for $\cchi=5$. This small contribution is seen to decrease strongly with the aspect ratio. This decrease may readily be predicted, assuming that the pressure on the ends scales as $\rho U^2\sim \rho\Omega^2L^2$. Since the end area is of $\mathcal{O}(D^2)$ and the magnitude of $\mathbf{x}\times\mathbf{n}$ is of $\mathcal{O}(D)$ there, the pressure contribution to the torque scales as $\rho\Omega^2L^2D^3$, which yields $C_{Tpe}\sim\cchi^{-2}$. Indeed, the behaviors reported in Fig. \ref{fig:ctmu_p_d}(b) are adequately fitted by the simple expression
\begin{equation}
C_{Tpe} =0.05\cchi ^{-2}\,.
\label{eq:Ctpd}
\end{equation}

\begin{figure}[h]
\centering
\includegraphics[width=6cm]{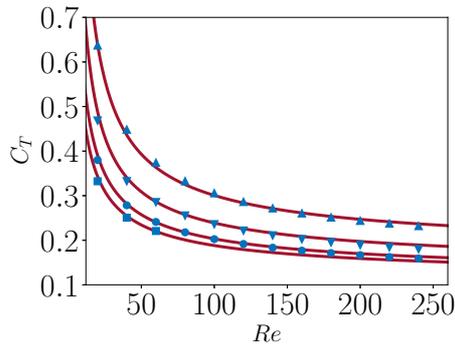} 
\caption{Torque coefficient as a function of $\text{Re}$ for cylinders with various aspect ratios. $\blacktriangle$: $\cchi =2$, $\blacktriangledown$: $\cchi =3$, $\bullet$: $\cchi =5$, $\blacksquare$: $\cchi =8$. Solid line: empirical fit (\ref{eq:Ct}).}
\label{fig:Ct}
\end{figure}
The total torque coefficient $C_T$ is eventually obtained by summing all the contributions fitted in (\ref{eq:Ctmul})-(\ref{eq:Ctpd}), keeping in mind that those of (\ref{eq:Ctmud}) and (\ref{eq:Ctpd}) have to be counted twice. This yields
\begin{equation}
C_{T} (\cchi,\text{Re})= 15 \cchi ^{-1.2}Re^{-1.2}+1.21\cchi ^{-0.23} Re ^{-0.75}+(1.32\cchi^{-2}+0.55)Re^{-1/2}+0.22\cchi ^{-2}+0.1\,.
\label{eq:Ct}
\end{equation}
As Fig. \ref{fig:Ct} shows, this empirical fit matches all numerical results well.



\section{Discussion}
\label{sec:discussion}

In this work, we used fully-resolved simulations to obtain approximate laws for the torque acting on a slender circular cylinder rotating about an axis passing through its centroid and perpendicular to its symmetry axis. For cylinders with an aspect ratio larger than $3$, we found the creeping-flow prediction based on the slender-body theory to agree well with numerical data, provided the expansion with respect to the small parameter $1/\ln(2\cchi)$ is carried out up to fourth order. We empirically modified the fourth-order term which we derived analytically, in such a way that the modified formula is valid down to $\cchi\approx2$. We carried out a series of runs in the range $0.1\leq\text{Re}\leq10$ to quantify finite-inertia effects. As is customary with slender bodies, numerical results revealed that the larger the aspect ratio the stronger the inertial increase of the torque for a given $\text{Re}$. Since no theoretical prediction is available for low-but-finite inertial effects in the configuration considered here, we merely introduced an empirical modification in the second-order term of the low-$\text{Re}$ prediction to account for these effects. The modified formula was found to fit the numerical data well up to $\text{Re}\approx10$ whatever the aspect ratio. Then we considered higher Reynolds numbers, up to $\text{Re}=240$. Numerical results revealed that the flow remains stationary throughout this range and preserves a planar symmetry with respect to the plane perpendicular to the rotation axis and passing through the body centroid. The flow exhibits a second symmetry, resulting from the combination of two mirror symmetries with respect to two mutually orthogonal planes. Because of this symmetry, the two cylinder ends provide identical contributions to the torque. We used numerical data to establish separate approximate fits for the contributions resulting from pressure and viscous effects on each part of the body surface, and provided simple physical arguments in support of the corresponding scaling laws. From a practical standpoint, the main outcomes of this investigation are formulas (\ref{eq:sl4_bism}), (\ref{eq:sl4_bism_re}) and (\ref{eq:Ct}) which accurately approach the rotation-induced torque from creeping-flow conditions up to $\text{Re}\approx250$ for all aspect ratios larger than $2$.\\
\indent These empirical laws allow the rotation rate of sedimenting cylindrical rod-like particles and fibers to be predicted in the various flow regimes. Consider first the low-but-finite Reynolds number regime. There, the orientation-induced inertial torque $T_o$ acting on a long cylinder translating with velocity $U$ in a fluid at rest is known to be of $\mathcal{O}(\mu U L^2 \cchi\text{Re}_\text{u} / (\ln \cchi) ^2)$ \citep{khayat1989}, with $\text{Re}_\text{u}$ the Reynolds number based on the body diameter and velocity $U$. In contrast, according to (\ref{eq:sl4_bism}), the rotation-induced torque scales as $T \sim \mu \Omega L^3 / \ln \cchi$, so that the ratio $T_o/T$ is of $\mathcal{O}( \frac{U}{\Omega L} \cchi\text{Re}_\text{u}/ \ln\cchi)$. When a rod-like particle with a uniform density distribution settles under the effect of gravity, its angular velocity must be such that $T$ and $T_o$ balance each other, assuming that conditions for a quasi-steady evolution are fulfilled (see \citep{cox1965,shin2006} for a discussion on this aspect). This quasi-steady balance implies
\begin{equation}
\frac{\Omega D}{U}\sim\text{Re}_\text{u}(\ln\cchi)^{-1}\,.
\label{scaling1}
\end{equation}
Similar to the approach followed in the previous section, the numerical results obtained in \cite{kharrouba2020} for $T_o$ in inertia-dominated regimes were synthesized in the form of an empirical fit. Considering that the cylinder axis makes an angle $\theta$ with the incoming velocity (only inclinations up to $30^\circ$ were considered in \cite{kharrouba2020}), and writing $T_o$ in the form $T_o(\cchi\gg1,\theta, \text{Re}_\text{u})\approx C_{To}(\text{Re}_\text{u}) \rho U^2L^2D\sin2\theta$, these results indicate that in the limit of large aspect ratios one has $C_{To}\approx0.69\cchi^{-0.47}\text{Re}_\text{u}^{-0.35}$. In the same limit, (\ref{eq:Ct}) reduces to $C_{T}\approx0.1+0.55\text{Re}^{-1/2}$. Equating the two torques and considering orientations such that $\sin2\theta\neq0$ yields for sufficiently large $\text{Re}$
\begin{equation}
 \frac{\Omega D}{U}\sim\text{Re}_\text{u}^{-0.175}\cchi^{-1.235}\,.
 \label{scaling2}
 \end{equation}
 However, the $\text{Re}^{-1/2}$-correction in (\ref{eq:Ct}) remains dominant up to $\text{Re}\approx30$. In the intermediate range, say $1\lesssim\text{Re}\lesssim20$, one then has
 \begin{equation}
 \frac{\Omega D}{U}\sim\text{Re}_\text{u}^{0.1}\cchi^{-1.31}\,.
 \label{scaling3}
 \end{equation}
 Comparing the above three predictions gives insight into the way the relative rotation rate $\Omega D/U$ varies with the translation Reynolds number and the body aspect ratio. Starting from the linear growth indicated by (\ref{scaling1}) for low-but-finite Reynolds numbers, (\ref{scaling3}) predicts that the growth with $\text{Re}_\text{u}$ slows down drastically for moderately inertial conditions, until $\Omega D/U$ eventually slowly decreases as $\text{Re}_\text{u}^{-0.175}$ under strongly inertial conditions. All predictions indicate that the larger the aspect ratio the slower the body rotation. However inertial effects sharpen the corresponding slowing down, since the $1/\ln\cchi$-variation typical of finite-slenderness effects at low $\text{Re}_\text{u}$, is replaced by a sharper $\cchi^{-n}$-variation with $n\gtrsim1$ for $\text{Re}_\text{u}>1$.






Throughout this work, we deliberately disregarded unsteady effects which are obviously significant, if not dominant, during collision processes \cite{gondret2002}. In such situations, the rotation-induced torque may be much larger than $T_o$,  the torque balance being then satisfied thanks to the time rate-of-change of the particle rotation rate. The collision frequency increases with the particle concentration, making these transient effects become increasingly important as denser suspensions are concerned. In such situations, it is also desirable to know the relaxation time beyond which the steady approximate formulas established here become valid again. However, the history force on a rotating slender body is not known in closed form in the Stokes regime. Similarly, in the inviscid limit, the inertia-induced (or added-mass) torque is not known in closed form, although approximate formulas have been proposed \citep{korotkin2009}. This lack of theoretical guides in the asymptotic regimes makes the development of approximate predictions for such unsteady situations challenging. However, given their practical relevance, we plan to consider these transient effects in future work.



\section*{Acknowledgments}
Mohammed Kharrouba's fellowship was provided by IFP Energies Nouvelles whose financial support is greatly appreciated. Part of the computations were carried out on the IRENE supercomputer under GENCI grant number A0072B10978.

\appendix
\section{Slender-body approximation for the rotation-induced torque}
\label{app:slender}
The torque on a long rotating body may be obtained through the slender-body theory \citep{batchelor1970,keller1976}. By expanding the solution in powers of the small parameter $\epsilon = 1 / \ln (2 \cchi)$, \citet{batchelor1970} determined the rotation-induced torque up to $\mathcal{O}(\epsilon^3)$. The coefficients of the corresponding expansion were obtained numerically. However, the logarithmic dependence of the loads with respect to $\cchi$ makes the expansion converge slowly, limiting the accuracy of the predictions for moderate aspect ratios. This is why including higher-order contributions is desirable. In this appendix, we derive the coefficients of the $\epsilon^n$-expansion up to $n=4$.\vspace{2mm}\\
\indent The total torque on a cylindrical body of length $L$ may be written as
\begin{equation}
\mathbf{T} = -8 \pi \mu L ^2 \int _0 ^1 \mathbf{x} \times \mathbf{f}(x) \mathrm{d} x\,,
\label{eq:tot}
\end{equation}

\noindent where $\mathbf{f}(x)$ is the density of the Stokeslet distribution along the body, $\mathbf{x}$ is the local position (with $\mathbf{x}=\mathbf{0}$ at the body centroid) and $x$ denotes the arc length. For a body rotating along an axis perpendicular to its symmetry axis (here along $z$ as in the main body of the paper), the previous expression reduces to
\begin{equation}
T = -8 \pi \mu L ^2 \int _0 ^1 x f_y(x) \mathrm{d} x\,.
\end{equation}
The Stokeslet density $f_y$ was obtained by \citet{keller1976}, using a matched asymptotic expansion technique. It may be expressed as 
\begin{equation}
\displaystyle f_y(x)=-\frac{\epsilon}{2}\left(U_y(x)+f_y(x)\left(\ln(4x(1-x))+1\right)+\int_{-x}^{1-x}\frac{f_y(x+t)-f_y(x)}{\vert t \vert}dt\right)\,,
\label{eq:slender_f_y}
\end{equation}

\noindent with $U_y$ the rotation-induced velocity of the cylinder in the $y$-direction. For $\cchi \gg 1$, $U_y =  \Omega L \left(x -1/2\right)$, which indicates that the rotation-induced flow is equivalent to a uniform shear flow of strength $\Omega$ centered at the body centroid. As detailed in \cite{keller1976}, $f_y(x)$ may be obtained in an iterative way. Setting first $f_y(x) = 0$ on the right-hand side of (\ref{eq:slender_f_y}), the first-order approximation is obtained as $\displaystyle f_y^{(1)}(x)=-\Omega L \left(x -1/2\right)\epsilon/2 $. Inserting the first-order solution in the right-hand side of (\ref{eq:slender_f_y}), the second-order correction is readily obtained as
\begin{equation}
f_y^{(2)}(x)=-\frac{\Omega L(x-1/2)}{2}\left(\epsilon-\frac{\epsilon ^2}{2}\left(\ln(4x(1-x))-1\right)\right)\,,
\end{equation}
since the integral term in (\ref{eq:slender_f_y}) reduces to $\int_{-x}^{1-x}\frac{t}{\vert t \vert} dt = 1-2x$. Following the same procedure, one obtains the third-order correction in the form 
\begin{eqnarray}
\nonumber
f_y^{(3)}(x)&=&-\frac{\Omega L(x-1/2)}{2}\left\{\epsilon-\frac{\epsilon ^2}{2}\left(\ln(4x(1-x))-1\right)\right. \\
&+&\left. \frac{\epsilon ^3}{4}\left[\left(\ln(4x(1-x))-1\right)\left(\ln(4x(1-x))+1\right) + \frac{g_y^{(3)}(x)}{x-1/2}\right]\right\} \,,
\end{eqnarray}
where $g_y ^{(3)}(x) = \int _{-x} ^{1-x}(h_y ^{(3)}(x+t)-h_y ^{(3)}(t))/|t|dt$, with $h_y ^{(3)}(x) = (x-1/2)\left(\ln(4x(1-x))-1\right)$. The latter integral may be evaluated analytically, yielding
%
\begin{align}
g_y^{(3)}(x) =& -3 + \frac{\pi ^2}{6} + 6x - \frac{\pi ^2}{3}x + 6\ln 2-12x\ln 2 +2\ln(1-x)-2x\ln(1-x)-2x\ln(x) \nonumber \\
              &+\left(\frac{1}{2} -x\right)\left(\text{Li}_2\left(\frac{x-1}{x}\right)+\text{Li}_2\left(\frac{x}{x-1}\right)\right)\,,
\end{align}
where $\text{Li}_2$ is the polylogarithm function with argument 2. At this stage, the present prediction may be compared with the coefficients computed numerically by \citet{batchelor1970}. For this purpose one needs the results 
\begin{eqnarray}
\nonumber
&&\int _0 ^1 x(x-1/2)\left(\ln(4x(1-x))-1\right)dx=-11/36+\ln 2/6\,,\\
\nonumber
 &&\int _0 ^1 x(x-1/2)\left(\ln(4x(1-x))-1\right)\left(\ln(4x(1-x))+1\right)dx=(95-3\pi^2+12\ln 2 (-8+\ln 8))/108\,\\
 \nonumber
&&\int _0 ^1xg_y ^{(3)}(x)dx =11/18 -\ln 2 /3\,.
\end{eqnarray}
The third-order approximation for the torque is then found to be
\begin{equation}
T^{(3)}=\frac{-\pi\mu\Omega L^3}{3}\left[\epsilon+\epsilon^2\left(\frac{11}{6}-\ln2\right)+\epsilon^3\left(\frac{161}{36}- \frac{\pi ^2}{12}-\frac{11}{3}\ln 2+(\ln 2)^2\right)\right].
\label{eq:Tsb3}
\end{equation}
The numerical evaluation of the second- and third-order terms in (\ref{eq:Tsb3}) agrees with the results provided in \cite{batchelor1970}. At next order, the Stokeslet density may be expressed as 
\begin{eqnarray}
\displaystyle f_y^{4}(x)&=& f_y^{(3)}(x)\\
\nonumber
&&+\frac{\Omega L(x-1/2)}{16}\epsilon^4\left\{\left(\ln(4x(1-x))+1\right)\left[\left(\ln(4x(1-x))-1\right)\left(\ln(4x(1-x))+1\right)+ \frac{g_y^{(3)}(x)}{x-1/2}\right]+ \frac{g_y^{(4)}(x)}{x-1/2}\right\}\,,
\end{eqnarray}
where $g_y ^{(4)}(x) = \int _{-x} ^{1-x}(h_y ^{(4)}(x+t)-h_y ^{(4)}(t))/|t|dt$ with $h_y ^{(4)}(x) = (x-1/2)\left(\ln(4x(1-x))-1\right)\left(\ln(4x(1-x))+1\right) + g_y^{(3)}(x)$. Integrating separately each contribution, one obtains
\begin{align}
\nonumber
&\int _0 ^1 x(x-1/2)\left(\ln(4x(1-x))-1\right)\left(\ln(4x(1-x))+1\right)^2dx = \nonumber \\
\nonumber
&\frac{1}{216} \left(216 \zeta (3)-1042+6 \ln 4 (85+3 \ln 4(\ln 4-7) )+\pi ^2 (42-9 \ln 16)\right),\\
\nonumber
&\int _0 ^1 xg_y^{(3)}\left(\ln(4x(1-x))+1\right)dx = \frac{1}{54} \left(-9 \zeta (3)-95+3 \pi ^2-36 \ln^2 (2)+96 \ln 2\right),\\
\nonumber
&\int _0 ^1 xg_y^{(4)}dx = \frac{1}{54} \left(-161+3 \pi ^2-9 \ln ^2(4)+54 \left(\ln 4+\frac{4}{9}\ln 2\right)\right).
\end{align}
Summing all contributions, the fourth-order approximation is eventually obtained as
\begin{equation}
T^{(4)}=T^{(3)} -\frac{\pi\mu\Omega L^3\epsilon^4}{3}\left[- \frac{5}{4}\zeta (3)+\frac{1033}{72}-\ln ^3 (2)+\frac{11}{2}\ln ^2(2) - \frac{161}{12}\ln 2-\pi ^2 \left(\frac{11}{24}-\frac{1}{4} \ln 2\right)\right]\,,                                            
\label{eq:sl4}
\end{equation}


                                           
\noindent where $\zeta$ denotes the Riemann Zeta function. Interestingly, at each order of the expansion one may remark that $\int _0 ^1 x\int_{-x}^{1-x}(f_y(x+t)-f_y(x))/\vert t \vert dt dx = -2\int_0^1{xf_y(x)dx}$. This statement may presumably be proved by mathematical induction, even though we did not attempt to do so.


\section{Minimum dissipation theorem for a rotating cylinder}
\label{app:min}

The minimum dissipation theorem states that in any geometrical configuration, the Stokes solution dissipates less energy than any other solenoidal solution satisfying the same boundary conditions. For a translating and rotating rigid body, the dissipation rate may be related to the rate of work of the force and torque. Therefore \citep{hill1956,kim2013}
\begin{equation}
\mathbf{F}\cdot\mathbf{U} + \mathbf{T} \cdot \boldsymbol{\Omega} \leq \Phi^*\,,
\label{eq:dissip}
\end{equation} 
where $\mathbf{F}$ and $\mathbf{T}$ are the force and torque acting on the body moving at velocity $\mathbf{U}$ and rotating with angular velocity $\boldsymbol{\Omega}$, and $\Phi^*$ is the dissipation rate of any other solenoidal velocity field $\mathbf{u^*}$ obeying the same boundary conditions. This theorem has been used successfully to estimate bounds for the drag acting on translating bodies \citep{hill1956}. The corollary for rotating bodies directly follows from (\ref{eq:dissip}). Assuming that a rigid body bounded by a surface $\mathcal{S}_i$ is immersed in a body of fluid bounded externally by a surface $\mathcal{S}_o$, (\ref{eq:dissip}) implies that the torque on $\mathcal{S}_i$ is smaller than that on $\mathcal{S}_o$. The proof of this statement is similar to the one provided in \cite{hill1956} for the drag force. The actual velocity field, say $\mathbf{u_i}$, is the one induced by the rotation of $\mathcal{S}_i$, whereas $\mathbf{u_o}$, the velocity field induced by the rotation of $\mathcal{S}_o$, may be used for $\mathbf{u^*}$. Assuming that the fluid located in between the two surfaces rotates as a solid with the velocity field $\boldsymbol{\Omega} \times \mathbf{x}$ implies that the associated dissipation rate is zero. This immediately yields $T_i \leq T_o$, where $T_i$ and $T_o$ denote the torques associated with the rotation of $\mathcal{S}_i$ and $\mathcal{S}_o$, respectively. Consider now a circular cylinder with $\cchi=1$. A sphere of diameter $D$ may be entirely enclosed within the cylinder, while a sphere of diameter $2^{1/2}D$ completely encloses it. It follows that the torque $T$ on the cylinder is such that $\pi \mu D^3 \Omega \leq \vert T\vert \leq 2^{3/2} \pi \mu D^3 \Omega $. 



\bibliographystyle{unsrtnat}
\bibliography{biblio}

\begin{thebibliography}{22}
\providecommand{\natexlab}[1]{#1}
\providecommand{\url}[1]{\texttt{#1}}
\expandafter\ifx\csname urlstyle\endcsname\relax
  \providecommand{\doi}[1]{doi: #1}\else
  \providecommand{\doi}{doi: \begingroup \urlstyle{rm}\Url}\fi

\bibitem[Batchelor(1970)]{batchelor1970}
G.~K. Batchelor.
\newblock Slender-body theory for particles of arbitrary cross-section in
  {S}tokes flow.
\newblock \emph{J. Fluid Mech.}, 44:\penalty0 419--440, 1970.

\bibitem[Khayat and Cox(1989)]{khayat1989}
R.~E. Khayat and R.~G. Cox.
\newblock Inertia effects on the motion of long slender bodies.
\newblock \emph{J. Fluid Mech.}, 209:\penalty0 435--462, 1989.

\bibitem[Subramanian and Koch(2005)]{subramanian2005}
G.~Subramanian and D.~L. Koch.
\newblock Inertial effects on fibre motion in simple shear flow.
\newblock \emph{J. Fluid Mech.}, 535:\penalty0 383--414, 2005.

\bibitem[Einarsson et~al.(2015)Einarsson, Candelier, Lundell, Angilella, and
  Mehlig]{einarsson2015}
J.~Einarsson, F.~Candelier, F.~Lundell, J-R. Angilella, and B.~Mehlig.
\newblock Effect of weak fluid inertia upon {J}effery orbits.
\newblock \emph{Phys. Rev. E}, 91:\penalty0 041002, 2015.

\bibitem[Cox(1965)]{cox1965}
R.~G. Cox.
\newblock The steady motion of a particle of arbitrary shape at small
  {R}eynolds numbers.
\newblock \emph{J. Fluid Mech.}, 23:\penalty0 625--643, 1965.

\bibitem[Lopez and Guazzelli(2017)]{lopez2017}
D.~Lopez and E.~Guazzelli.
\newblock Inertial effects on fibers settling in a vortical flow.
\newblock \emph{Phys. Rev. Fluids}, 2:\penalty0 024306, 2017.

\bibitem[Roy et~al.(2019)Roy, Hamati, Tierney, Koch, and Voth]{roy2019}
A.~Roy, R.~J. Hamati, L.~Tierney, D.~L. Koch, and G.~A. Voth.
\newblock Inertial torques and a symmetry breaking orientational transition in
  the sedimentation of slender fibres.
\newblock \emph{J. Fluid Mech.}, 875:\penalty0 576--596, 2019.

\bibitem[Dennis et~al.(1980)Dennis, Singh, and Ingham]{dennis1980}
S.~C.~R. Dennis, S.~N. Singh, and D.~B. Ingham.
\newblock The steady flow due to a rotating sphere at low and moderate
  {R}eynolds numbers.
\newblock \emph{J. Fluid Mech.}, 101:\penalty0 257--279, 1980.

\bibitem[Kirchhoff(1876)]{kirchhoff1876}
G.~Kirchhoff.
\newblock \emph{Vorlesungen \"uber Mathematische Physik. Mechanik}.
\newblock Teubner, 1876.

\bibitem[Loth(2008)]{loth2008}
E.~Loth.
\newblock Drag of non-spherical solid particles of regular and irregular shape.
\newblock \emph{Powder Technol.}, 182:\penalty0 342--353, 2008.

\bibitem[Kry and List(1974{\natexlab{a}})]{kry1974}
P.~R. Kry and R.~List.
\newblock Aerodynamic torques on rotating oblate spheroids.
\newblock \emph{Phys. Fluids}, 17:\penalty0 1087--1092, 1974{\natexlab{a}}.

\bibitem[Kry and List(1974{\natexlab{b}})]{kry1974b}
P.~R. Kry and R.~List.
\newblock Angular motions of freely falling spheroidal hailstone models.
\newblock \emph{Phys. Fluids}, 17:\penalty0 1093--1102, 1974{\natexlab{b}}.

\bibitem[Kharrouba et~al.(2021)Kharrouba, Pierson, and
  Magnaudet]{kharrouba2020}
M.~Kharrouba, J-L. Pierson, and J.~Magnaudet.
\newblock Flow structure and loads over inclined cylindrical rodlike particles
  and fibers.
\newblock \emph{Phys. Rev. Fluids}, under revision, 2021.

\bibitem[Mougin and Magnaudet(2002)]{mougin2002}
G.~Mougin and J.~Magnaudet.
\newblock The generalized {K}irchhoff equations and their application to the
  interaction between a rigid body and an arbitrary time-dependent viscous
  flow.
\newblock \emph{Int. J. Multiphase Flow}, 28:\penalty0 1837--1851, 2002.

\bibitem[Magnaudet et~al.(1995)Magnaudet, Rivero, and Fabre]{magnaudet1995}
J.~Magnaudet, M.~Rivero, and J.~Fabre.
\newblock Accelerated flows past a rigid sphere or a spherical bubble. {P}art
  1. {S}teady straining flow.
\newblock \emph{J. Fluid Mech.}, 284:\penalty0 97--135, 1995.

\bibitem[Keller and Rubinow(1976)]{keller1976}
J.~B. Keller and S.~I. Rubinow.
\newblock Slender-body theory for slow viscous flow.
\newblock \emph{J. Fluid Mech.}, 75:\penalty0 705--714, 1976.

\bibitem[Hunt et~al.(1988)Hunt, Wray, and Moin]{hunt1988}
J.~C.~R. Hunt, A.~A. Wray, and P.~Moin.
\newblock Eddies, streams, and convergence zones in turbulent flows.
\newblock Center for Turbulence Research Report CTR-S88, 1988.

\bibitem[Shin et~al.(2006)Shin, Koch, and Subramanian]{shin2006}
M.~Shin, D.~L. Koch, and G.~Subramanian.
\newblock A pseudospectral method to evaluate the fluid velocity produced by an
  array of translating slender fibers.
\newblock \emph{Phys. Fluids}, 18:\penalty0 063301, 2006.

\bibitem[Gondret et~al.(2002)Gondret, Lance, and Petit]{gondret2002}
P.~Gondret, M.~Lance, and L.~Petit.
\newblock Bouncing motion of spherical particles in fluids.
\newblock \emph{Phys. Fluids}, 14:\penalty0 643--652, 2002.

\bibitem[Korotkin(2009)]{korotkin2009}
A.~I. Korotkin.
\newblock \emph{Added Masses of Ship Structures}.
\newblock Springer, 2009.

\bibitem[Hill and Power(1956)]{hill1956}
R.~Hill and G.~Power.
\newblock Extremum principles for slow viscous flow and the approximate
  calculation of drag.
\newblock \emph{Q. J. Mech. Appl. Math.}, 9:\penalty0 313--319, 1956.

\bibitem[Kim and Karrila(1991)]{kim2013}
S.~Kim and S.~J. Karrila.
\newblock \emph{Microhydrodynamics: Principles and Selected Applications}.
\newblock Butterworth-Heinemann, 1991.

\end{thebibliography}
\end{document}